\documentclass[aip,jcp,reprint]{revtex4-1}
\usepackage{graphicx}
\usepackage{amsmath,amssymb}

\begin{document}

\title{Interaction between random heterogeneously charged surfaces in an electrolyte solution}

\author{Amin Bakhshandeh}
\email{bakhshandeh.amin@gmail.com}
\affiliation{Departamento de F\'\i sica, Instituto de F\'\i sica e Matem\'atica, Universidade Federal de Pelotas,
Caixa Postal 354, CEP 96010-900, Pelotas, RS, Brazil}

\author{Alexandre P. dos Santos}
\email{alexandre.pereira@ufrgs.br}
\affiliation{Instituto de F\'\i sica, Universidade Federal do Rio Grande do Sul, Caixa Postal 15051, CEP 91501-970,
Porto Alegre, RS, Brazil}

\author{Alexandre Diehl}
\email{diehl@ufpel.edu.br}
\affiliation{Departamento de F\'\i sica, Instituto de F\'\i sica e Matem\'atica, Universidade Federal de Pelotas,
Caixa Postal 354, CEP 96010-900, Pelotas, RS, Brazil}

\author{Yan Levin}
\email{levin@if.ufrgs.br}
\affiliation{Instituto de F\'\i sica, Universidade Federal do Rio Grande do Sul, Caixa Postal 15051, CEP 91501-970,
Porto Alegre, RS, Brazil}

\date{\today}

\begin{abstract}
We study, using Monte Carlo simulations, 
the interaction between infinite heterogeneously charged surfaces inside an electrolyte solution. 
The surfaces are overall neutral
with quenched charged domains.  An average over the quenched disorder is performed to obtain the net force. We find that
the interaction between the surfaces is repulsive at short distances and is attractive for larger separations.

\end{abstract}

\maketitle

\section{Introduction}

In physical chemistry and biophysics one often finds situations in which electrolyte solution is confined between charged surfaces.
The surfaces can belong to macromolecules, colloidal particles, electrodes or membranes.  Presence of electrolyte between surfaces 
can strongly modify the interaction between them.~\cite{Israelachvili,Evans}.

Over seventy years ago, Derjaguin,
Landau, Verwey and Overbeek (DLVO) presented a theory which accounts for the interaction between weakly charged homogeneous surfaces~\cite{DeLa41,VeOv48}. 
The net interaction between two surfaces was attributed to the 
electrostatic double layer forces and the van der Waals force. The van der Waals force dominates when the separation between the surfaces is small, while the electrostatic repulsion is dominant on larger length scales. The theory works reasonably well for weakly charged homogeneously surfaces~\cite{Levin} and has been widely used to study colloidal stability. It fails, however, to account for the correlation induced attraction between like-charged objects inside an electrolyte solution containing multivalent counterions~\cite{Pa80,GuJo84,StRo90,LaGr97,SoDe99,LeAr99,LiLo99,BuAn03,SaTr11,Zh13} or for ionic specificity~\cite{BoWi01,LoJo03,LoSa08,PeOr10,DoLe11}.

It is also well known that two surfaces with annealed positive and negatively charged domains feel attraction~\cite{MiCh94,Mi95,ZhYo05,MeLi05,BeBu07}. In this case, positive domains on one surface become correlated with the negatively charged domains on another surface, resulting in an attractive interaction~\cite{Tsao,Brewster}.  Jho~{\it et al.}~\cite{Jho} carried out numerical simulations for flat surfaces with movable charged domains.  Long-range attractive force was observed and the mechanism behind the attraction was found to be the positional  correlation between oppositely charged domains. 

Recently, Silbert {\it et al.}~\cite{Silbert} conducted an interesting experiment to explore the interaction between
heterogeneously charged surfaces. Remarkably, they observed an attraction which extended up to 500 \AA.  At first the attraction was attributed to the correlation between the oppositely charged domains on the two surfaces.   However when a rapid shear motion was introduced between the surfaces to frustrate the correlations, the attraction persisted. Since the charged domains under the shear motion could not become correlated, the distribution of surface
charge was effectively quenched. It became clear that the correlations weren't the mechanism responsible for the  attraction in the experiments of Silbert {\it et al.}. The question then became: what was?  Silbert {\it et al.} attributed the attraction to the unequal nature of the repulsive and attractive interactions between like-charged and oppositely-charged domains. This, however, appeared to contradic the conclusion that
quenched charge disorder should not lead to attraction between heterogeneously charged surfaces~\cite{Naji, Andelman, Sarabadani}.  To justify their conclusion,  Silbert {\it et al.} presented a simple argument based on the Poisson-Boltzmann~(PB) equation. They suggested that the interaction between two neutral surfaces with a quenched charge 
disorder arises from an asymmetry in the interaction between like and oppositely-charged domains inside an electrolyte solution.  For like-charged domains, the counterions are required to stay between the surfaces to preserve the local charge neutrality, while for the oppositely charged domains this is not necessary.  The entropic contribution to the overall force is, therefore, asymmetric in the two cases.  Silbert {\it et al.}, then, suggested that the force between the two heterogeneously
charged random surfaces can be estimated as an arithmetic average of the force between two like-charged and two-oppositely charged homogeneous surfaces.

In the present work, we will use Monte Carlo simulations to show that the conclusions of Silbert {\it et al.} are qualitatively correct.  
The long-range attraction between 
heterogeneous surfaces with a quenched disorder arises due to the asymmetric interaction between oppositely and like-charged domains. On the
other hand, we will demonstrate that a simple arithmetic average of the force between like-charged and oppositely-charged homogeneous 
surfaces is not sufficient to quantitatively account for the range and strength of the attraction between heterogeneous surfaces with a quenched charge disorder,  and a more sophisticated calculation must be performed.  The paper is organized as follows. In section~\ref{model}, we explain the model and the simulation details. In section~\ref{results}, we summarize our results. In section~\ref{conclusions} we conclude our work.

\section{THE MODEL AND SIMULATION DETAILS}\label{model}

\begin{figure}[h]
\includegraphics[width=8cm]{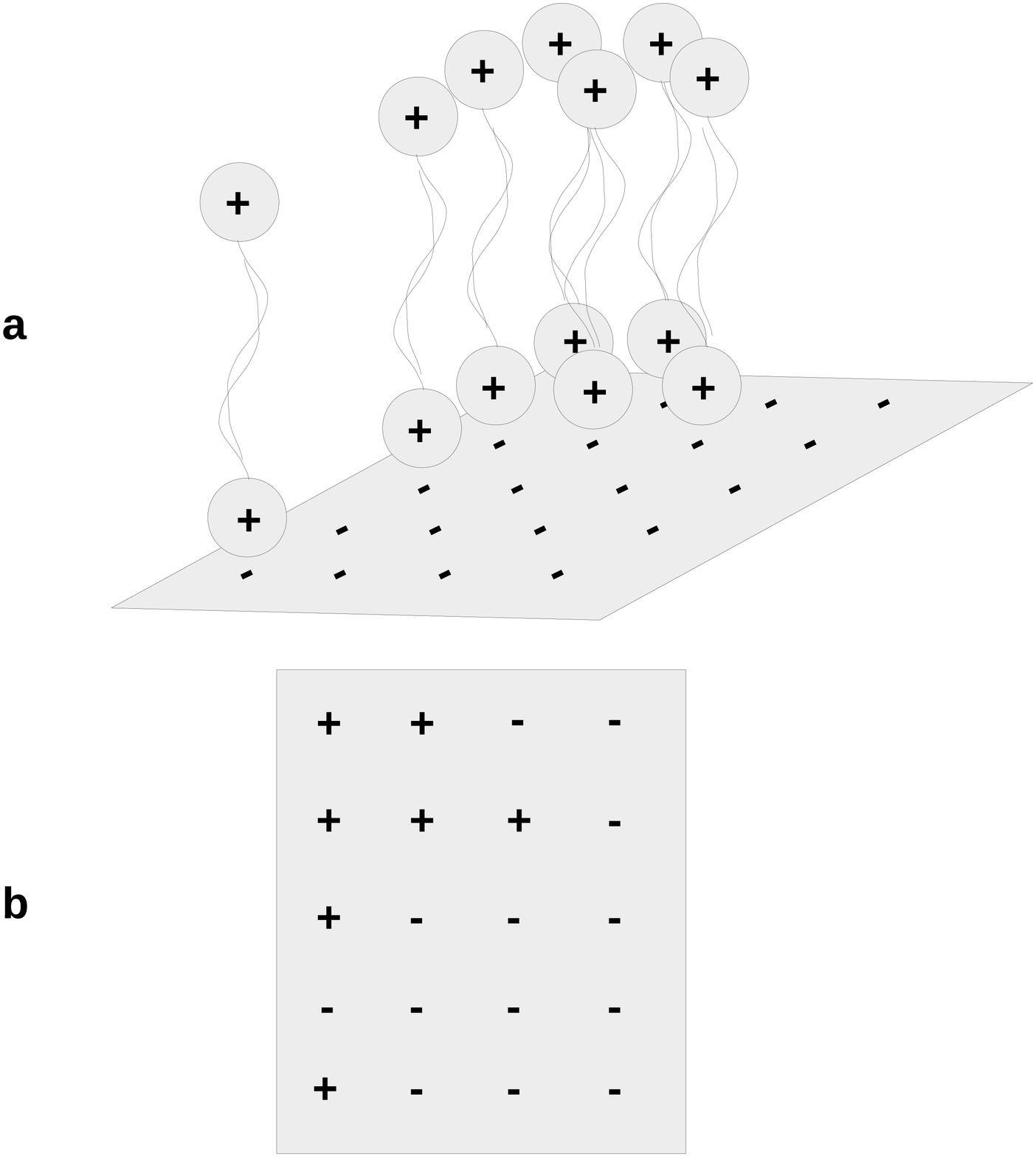}
\caption{Representation of the domains. Negative sites with absorbed surfactants can be considered as positive sites. In ``a" a side view and in ``b" a top view of the wall.}
\label{f1}
\end{figure}

In the experiments of Silbert {\it et al.} the charged domains were produced by the adsorption of cationic micelles to an anionic substrate.  To simplify the calculations we will neglect the spatial extent of micelles and project all the charge onto a flat surface, see Fig.~\ref{f1}. As an additional simplification, we will  divide our surfaces into positive and negative domains of the same surface area.

Our system then consists of two flat surfaces of dimensions $L_x$ and $L_y$, enclosing an electrolyte solution. For simplicity we set $L_x=L_y=400$  \AA. The plates are separated by a distance $L$.  The solvent is assumed to be a uniform dielectric of permittivity $\epsilon_w$. 
The Bjerrum length, defined as $\lambda_B = q^2/k_BT\epsilon_w$ --- where $q$, $k_B$ and $T$ are the proton charge, the Boltzmann constant, and the temperature, respectively  ---
is $7.2~$\AA $\,$ for water at room temperature.    The ions are modeled as hard  spheres of radius $2~$\AA. To perform simulations we use a grand canonical Monte Carlo algorithm~(GCMC)~\cite{Valleau,Frenkel,Allen}, see appendix~\ref{appen1} for details.  The system is in contact with a salt reservoir at concentration $\rho_s$. As an input, the GCMC requires the chemical potential of the 
ions of reservoir.  For 1:1 electrolyte this can be calculated from $\rho_s$ using the mean spherical approximation~(MSA)~\cite{Ching,Fisher,LeWa72a,LeWa72b}, which is very
accurate for weakly interacting ions.  Similarly, MSA also provides us with the osmotic pressure of the bulk electrolyte.  The force per unit area between the two surfaces is then the pressure between the two plates, minus the pressure of the bulk (reservoir) electrolyte.  
The pressure on each plate is calculated by taking into account the electrostatic interactions and the entropic force arising from the momentum transfer during the collisions of the ions with the surfaces.  The entropic contribution is calculated using the method of Wu~{\it et al.}~\cite{Wu,Alexander}. The details of the calculation are presented in appendix~\ref{appen3}.

To obtain the chemical potential and the osmotic pressure of a reservoir containing  2:1 electrolyte we first perform a bulk GCMC simulation.
In this simulation the chemical potential of electrolyte is fixed and the average concentration of ions inside the reservoir is calculated.  We then perform a NPT simulation to calculate the osmotic pressure of the electrolyte~\cite{Allen} at this concentration. The NPT MC simulation method is described in the appendix~\ref{appen2}. To calculate the electrostatic energy, a 3D Ewald summation method with a correction 
for the slab geometry of Yeh and Berkowitz~\cite{Yeh} is used.


\begin{figure}[t]
\includegraphics[width=8cm]{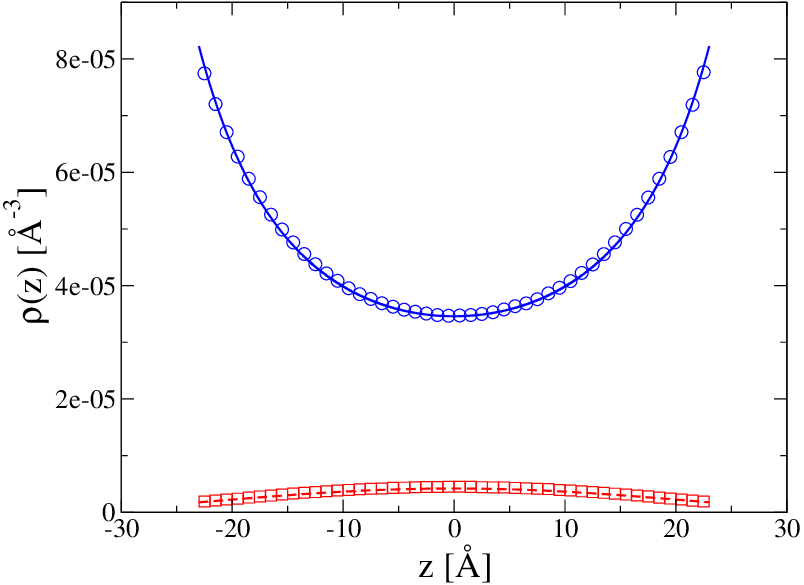}
\caption{Ionic density profiles between two equaly charged plates, with charge density $-0.01602~$C/m$^2$. Symbols represent simulations data while lines represent PB curves. The concentration of the monovalent salt in reservoir is $20~$mM. The solid line and circles represent positive ions while the dashed line and squares, negative ones. $z$ is the position between two surfaces.}
\label{f4}
\end{figure}
\begin{figure}[t]
\includegraphics[width=8cm]{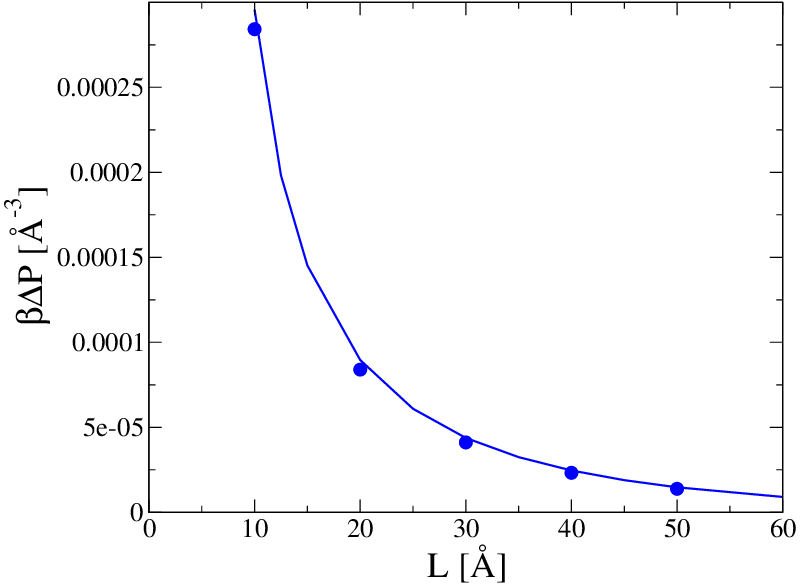}
\caption{Osmotic pressure for two equaly charged plates. The parameters are the same as Fig.~\ref{f4}.}
\label{f5}
\end{figure}

\begin{figure}[t]
\includegraphics[width=8cm]{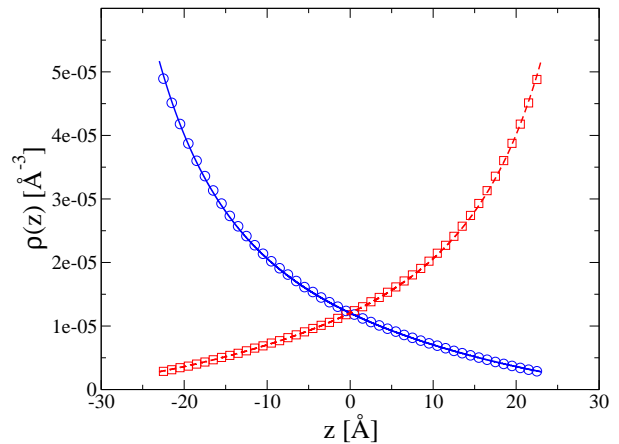}
\caption{Ionic density profiles between two oppositely charged plates, with charge density $\pm 0.01602~$C/m$^2$. Symbols represent simulations data while lines represent PB curves. The concentration of the monovalent salt in reservoir is $20~$mM. The solid line and circles represent positive ions while the dashed line and squares, negative ones. $z$ is the position between two surfaces.}
\label{f6}
\end{figure}

\begin{figure}[h]
\includegraphics[width=8cm]{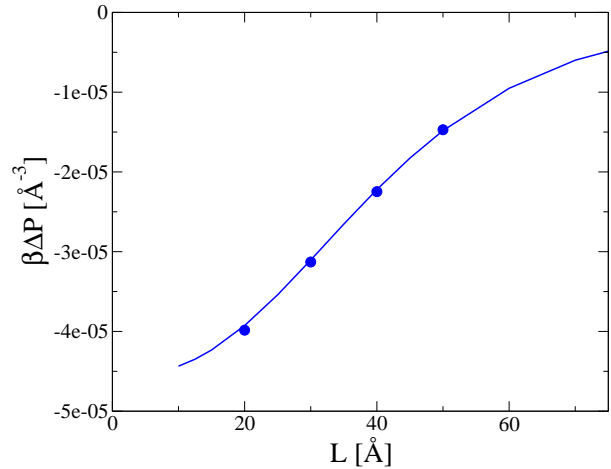}
\caption{Osmotic pressure for two oppositely charged plates. The parameters are the same as Fig.~\ref{f6}.}
\label{f7}
\end{figure}

Before considering the interaction between heterogeneously charged surfaces, we investigate two simpler cases: equally-charged and oppositely-charged homeogeneous surfaces, both in contact with a monovalent salt reservoir. In this case each plate is formed by $N_s^2$ point charge pseudo-particles, uniformly distributed, with separation $L_x/N_s$ along the surface.   The charge of the pseudo particles is adjusted to obtain the desired surface charge density.  The system has periodic boundary condition in $x$ and $y$ directions.  We set $N_s=40$. To test the simulations, we compare our results with the solution of the Poisson-Boltzmann (PB) equation. For weakly charged homogeneous surfaces inside a dilute 1:1 electrolyte PB equation is expected to be very 
accurate~\cite{Levin}.   The algorithm to solve the PB is the same as in Ref.~\cite{TaLe98} adapted to the slab geometry. In Fig.~\ref{f4}, we compare the density profiles obtained using the simulations to the solution of PB equation. As expected, a very good agreement between simulations and theory is obtained.  For homogeneously charged plates the force between the surfaces can be easily calculated using the contact theorem~\cite{HeBl79,BlHe81,CaCh81}.  The net force per unit area is the difference between internal and external pressures.   As mentioned above, the osmotic pressure of the reservoir can be obtained using MSA, which agrees perfectly with the NPT simulations.   
On the other hand, within the PB approximation the electrostatic correlations between the ions are completely ignored, and the bulk pressure of 1:1 electrolyte is simply that of an ideal gas.  In spite of this very crude approximation, for the parameters considered we see an excellent agreement between the simulations and theory, see Fig.~\ref{f5}. In Figs.~\ref{f6} and \ref{f7}, we show
the ionic distribution and the force per unit area for two oppositely charged homogeneous surfaces.  Again the agreement between GCMC simulations and PB equation is very good. 

To calculate the force  between two randomly charged heterogeneous surfaces we divided each plate into equi-sized domains, half of which
are positively charged while the other half are negatively charged. For each charge distribution,  we calculate 
the force between the two surfaces.  The net force is then calculated as an arithmetic average over the quenched disorder.
Because the number of configurations grows exponentially fast with the number of charged domains, in this paper we will consider 
surfaces with only two and four regions.  

\begin{figure}[h]
\includegraphics[width=5cm]{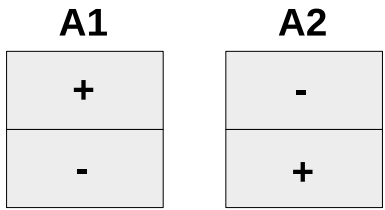}
\caption{Representation of  surfaces when they are divided into two charged domains. The surfaces are overall neutral.}
\label{f2}
\end{figure}

\begin{figure}[h]
\includegraphics[width=7cm]{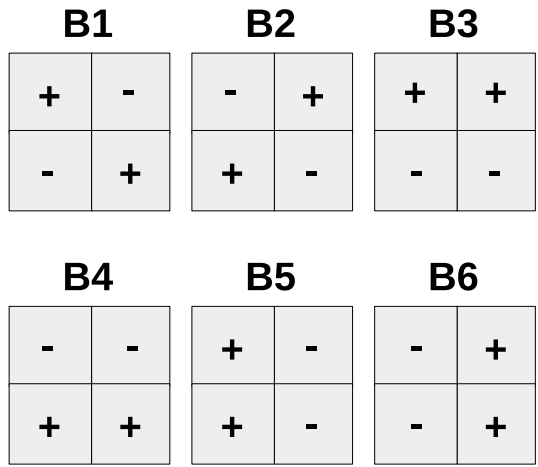}
\caption{Representation of  surfaces when they are divided into four charged domains. The surfaces are overall neutral.}
\label{f3}
\end{figure}

We start by considering two overall neutral plates with two charged domains each: one positive and one negative, as illustrated in Fig.~\ref{f2}.  There are two possible configurations, A1-A1 and A1-A2. The net force is the average over these two configuration. 
We next consider the interaction between two plates containing 4 charged regions, two positive and two negative.   This is illustrated in Fig.~\ref{f3}. There are $6$ possibilities for each plate. As a result, for two interacting plates, there are $36$ distinct configurations. However, various of these configurations are degenerate and are connected by symmetry.  We calculated the energy for each configuration and obtained the degeneracy factors which are presented in Tab.~\ref{table:nonlin}. All parameters for this model are the same as in the previous case. By dividing the plate into four areas, the positive and negative domains become smaller, in comparison with the two-region model. This way we are able to explore the effect of domain size on the overall interaction between the surfaces.  Furthermore, note that because of the periodic boundary conditions imposed by the Ewald summation, the plates are actually infinite, with a charge distribution corresponding to stripes or a checkerboard  patterns.    
\begin{table}[h]
\begin{tabular}{c c c c c c c}
\hline
Configurations: & B1-B1 & B1-B2 & B1-B3 & B1-B5 & B1-B4 & B3-B3 \\ 
\hline
Degeneracies: & 2 & 2 & 16 & 8 & 4 & 4 \\
\hline
\end{tabular}
\caption{Number of different configurations for the four-region model.}
\label{table:nonlin}
\end{table}

\section{Discussion}\label{results}

\subsection{Monovalent salt ions}

\begin{figure}
\includegraphics[width=8cm]{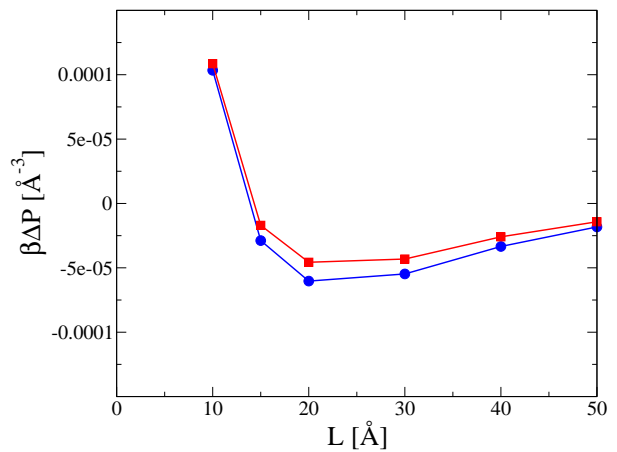}
\caption{Net force per unit area between two neutral surfaces with absolute site charge density equal to $0.056~$C/m$^2$, in contact with a salt reservoir
at concentration $0.01~$M. The circles and squares represent two and four-region models, respectively. The lines are guides to the eye.}
\label{f8}
\end{figure}

In the two-region model, the plates are divided into two domains, one positive and one negative, as shown in Fig.~\ref{f2}. The surface charge densities of both domains are the same. In Fig.~\ref{f8}, we plot the average (over disorder) force per unit area  between the two plates. For configuration A1-A1, at all distances repulsion is observed.  In configuration A1-A2, the two plates feel attraction.  In this case the number of ions between the plates is smaller, and the entropic force is reduced. Calculating the average over the two charge distributions, we see that the net force is repulsive at short distances, but becomes attractive at larger separations,  see in Fig.~\ref{f8}.
It should be mentioned that in all cases the electrostatic contribution to the force is attractive, as is expected for a charge neutral system. The net repulsion at short distances is observed because of a strong 
entropic force produced by the confined counterions.

\begin{figure}
\includegraphics[width=8cm]{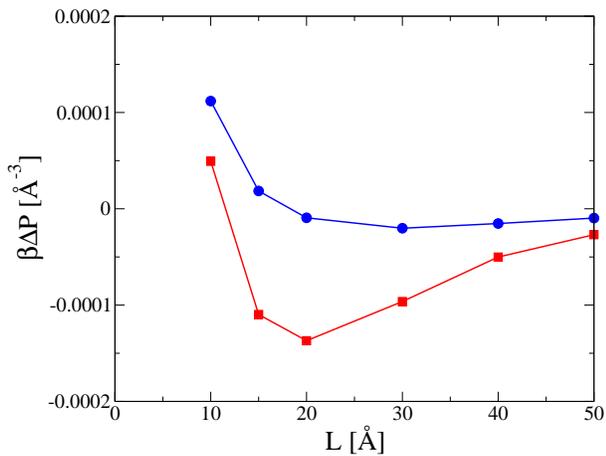}
\caption{Net force per unit area between surfaces with two charged region. The reservoir 1:1 salt concentration is $0.01~$M. The circles and squares represent the charge densities $0.04~$C/m$^2$ and $0.072~$C/m$^2$, respectively. The lines are guides to the eye.}
\label{f9}
\end{figure}

In Fig.~\ref{f9}, we plot the net force between two randomly charged surfaces for various surface charge densities of the two domains. As can be seen, by reducing the charge density, the attraction between the plates decreases. This is not surprising since the attraction 
is caused by the electrostatic interaction.

\begin{figure}[b]
\includegraphics[width=8cm]{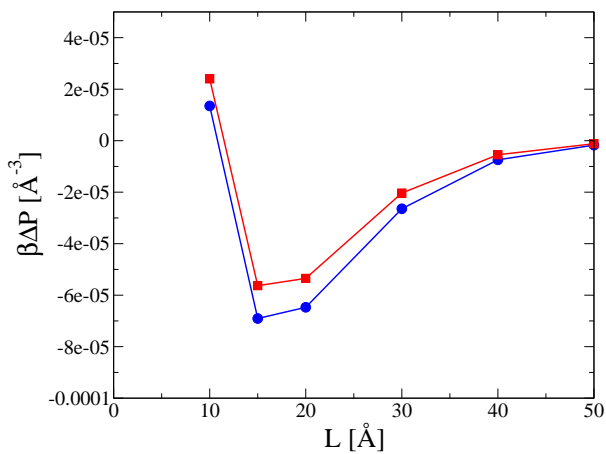}
\caption{A net force per unit area between two neutral surfaces with absolute site charge density equal to $0.056~$C/m$^2$, for 2:1 salt reservoir at concentration $10~$mM. The circles and squares represent two and four sites model, respectively. The lines are guides to the eye.}
\label{f10}
\end{figure}
To investigate the effect of domain size on the interaction between the surfaces, we divided the area of each plate into four regions. By doing this, each charged domain becomes smaller. There are a total of 36 different configurations many of which, however, are related by symmetry. We find that there are only 6 distinct arrangements, each with its own degeneracy factor listed in
Tab.~\ref{table:nonlin}.
The net force can now be easily calculated as a weighed average over the configurations listed in Tab.~\ref{table:nonlin}. Similar to the
 two-region model, a net attraction is observed, see Fig. \ref{f8}. 
We see, however, that the attraction is somewhat weaker than for the two-region model. 
 
\subsection{Divalent salt ions}

\begin{figure}
\includegraphics[width=8cm]{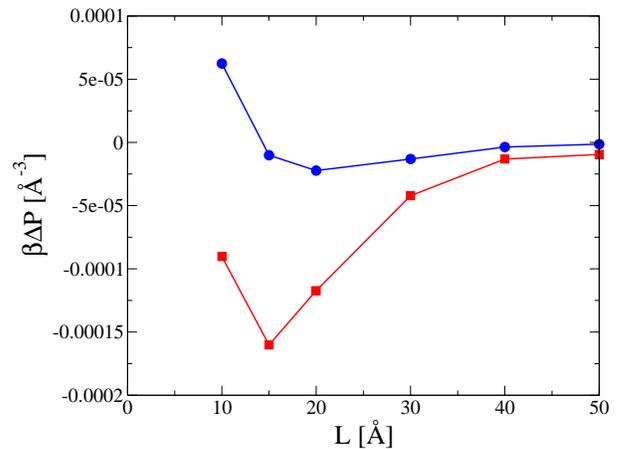}
\caption{Net force for various surface charge densities of the domains. The two-region model is used. The reservoir 2:1 salt concentration is $0.01~$M. The circles and squares represent the charge densities $0.04~$C/m$^2$ and $0.072~$C/m$^2$, respectively. The lines are guides to the eye.}
\label{f11}
\end{figure}

We next explore the effect of the charge asymmetry of electrolyte on the interaction between two heterogeneously charged surfaces.  We consider a reservoir of 2:1 electrolyte at concentration $10~$mM. The surface patterns and surface charge densities are the same as for the monovalent salt. Fig.~\ref{f10} shows that a charge asymmetric electrolyte leads to a stronger attraction between two
heterogeneous  random surfaces than a 1:1 electrolyte.   Furthermore,  in the case of 2:1 salt, the difference between the force for two and four-region models is larger, see Fig.~\ref{f10}, showing that the size of the charged domains is more important for the  interaction between the surfaces in asymmetric electrolyte solutions.
In Fig.~\ref{f11}, we plot the net force  for various surface charge densities.  Similar to what was observed for 1:1 electrolyte, decreasing the surface charge density of the domains diminishes the attraction between the plates.
\begin{figure}
\includegraphics[width=8cm]{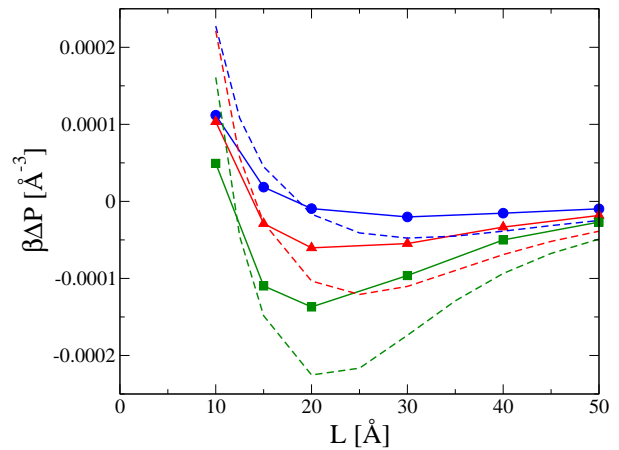}
\caption{Comparison of the calculation of Silber et al.~\cite{Silbert} with the results of the GCMC simulation for the 
two-region model for various charge densities of the domains, from top to bottom $0.04$, $0,056$ and  $0.072 C/m^2$.  
The calculation of Silber et al. significantly overestimates the attraction.  The deviations grow with the increasing 
surface charge density of the patches. Solid lines are the results of GCMC while the dashed lines are the results of the 
model of Silber et al.}
\label{f12}
\end{figure}
\section{Conclusion}\label{conclusions}
We have presented a simple model of interaction between randomly charged 
heterogeneous surfaces inside an electrolyte solution.  
Two and four-region models were explored.  Unfortunately due to the exponential growth of configurations, a study of surfaces with smaller charge domains is not viable.  For example a surface with $16$ domains would lead to a total of  $165636900$ different configurations,
a brute force study of which is clearly impossible.  Nevertheless, exploration of two and 4 region models has already provided a number of valuable insights.  Indeed as was argued by Silbert {\it et al.}, interaction between two overall neutral surfaces with randomly charged domains 
is attractive at large distances.  We find that the attraction decreases with the size of domains and increases with the charge asymmetry of the electrolyte solution. Silber {\it et al.} attributed the attraction between neutral randomly charged surfaces to the asymmetry of the interaction between like and oppositely-charged domains.  While the interaction between like-charged domains has a strong entropic component, the entropy is less important for oppositely charged domains, since the counterions are not required to stay between the oppositely charged regions to neutralize their charge. Based on this observation, Silber {\it et al.} concluded that the interaction between 
two heterogeneous random surfaces can be estimated as an arithmetic average of the force between two infinitely large like-charged surfaces and the force between two oppositely-charged surfaces.  As we saw in Section \ref{model}, the interaction between like-charged and  oppositely charged surfaces can be very accurately calculated using the PB theory.  We can, therefore, easily check the model of Silber {\it et al.}
by comparing it with our simulations.  In Fig.~\ref{f12} we contrast the model of Silber {\it et al.} with the results of our GCMC 
simulations for the two-region model.  Although qualitatively correct, the calculation of Silber et al. significantly overestimates the
attraction between the two surfaces.  The error increases with the increasing surface charge density of the charged domains.  
Finally, we expect that in the limit that the area of domains becomes small, the approach of Naji and Podgornik for randomly charged surfaces should become accurate~\cite{Naji}. Unfortunately, we are not able to check this limit in our model, since, as mentioned previously,  reduction of the domain size leads to an exponential growth of configurations.

Up to now we have only considered overall neutral surfaces.  In the future it will be interesting two explore the effect of breaking the charge neutrality on the attraction between heterogeneous randomly charged objects.
\section{Acknowledgments}
This work was partially supported by the CNPq, CAPES, INCT-FCx, and by the US-AFOSR under the grant 
FA9550-12-1-0438.

\begin{appendix}

\section{Grand Canonical Monte Carlo Simulation for $\alpha$:$1$ salt}\label{appen1}

Here, we briefly review the GCMC method for $\alpha$:$1$ salt. In each step, we have three possibilities, simple movement of ions and the addition or removal of ions. In order to keep the charge neutrality, if one randomly adds or removes a cation with valency $\alpha$, $\alpha$ anions must also be added or removed. The probability of a particular state $i$ is proportional to~\cite{Allen} 
\begin{equation}
\rho_i = \frac{V^{(N_+ + N_-)}e^{-\beta E_i + \beta N_+ \mu_+ + \beta N_- \mu_-}}{N_+!N_-!\Lambda_+^{3 N_+}\Lambda_-^{3 N_-}} \ ,
\end{equation}
where $V$ is the volume accessible to particles, $N_\pm$ are the number of cations and anions, $E_i$ is the electrostatic energy of the state, $\mu_\pm$ are the chemical potentials and $\Lambda_\pm$ are the thermal de Broglie wavelengths.

The transition probability for addition is:
\begin{equation}
\frac{\rho_j}{\rho_i} = \frac{V^{\alpha+1}e^{-\beta E_j+\beta E_i + \beta\mu_+ + \alpha\beta\mu_-}}{(N_+ + 1)(N_-+\alpha)(N_-+\alpha-1)...(N_-+1)\Lambda_+^{3}\Lambda_-^{3\alpha}} \ ,
\end{equation}
where $j$ is the state after addition. We define the parameter $z=e^{\beta\mu_+ + \alpha\beta\mu_-}/\Lambda_+^{3}\Lambda_-^{3\alpha}$. This parameter is the input of simulations. The transition probability for addition can be rewritten as
\begin{equation}
\frac{\rho_j}{\rho_i} = \frac{z\ V^{\alpha+1}e^{-\beta E_j+\beta E_i}}{(N_+ + 1)(N_-+\alpha)(N_-+\alpha-1)...(N_-+1)} \ .
\end{equation}

The thansition probability for the removal is:
\begin{equation}
\frac{\rho_j}{\rho_i} = \frac{e^{-\beta E_j+\beta E_i}N_+ N_-(N_- -1)...(N_--\alpha+1)}{z\ V^{\alpha+1}} \ ,
\end{equation}
where $j$ is the state after removal. After the transition probabilities have been calculated, they are compared with a random number, uniformly distributed between $0$ and $1$. If this random number is lower than the transition probability, the movement is accepted. Otherwise it is rejected.

\section{Entropic force}\label{appen3}

The algorithm of  Wu~{\it et al.}~\cite{Wu}, constructed to calculate the entropic force between two colloidal particles, can be easily adapted to the planar geometry.  The entropic force is give by the expression
\begin{equation}
\beta F = \frac{\left< N_1 \right>}{\Delta z} \ ,
\end{equation}
where $N_1$ is the number of overlaps of one of the walls with the free ions (which are held fixed)  after a displacement $\Delta z$. 
The force is obtained in the limit $\Delta z \rightarrow 0$. The entropic pressure is
\begin{equation}
\beta P = \frac{\left< N_1 \right>}{\Delta z L_x L_y} \ .
\end{equation}

\section{NPT Monte Carlo simulations}\label{appen2}

In this appendix, the NPT MC simulation is rapidly reviewed. Besides the particle movement, the volume of the simulation box is varied. This kind of movement is randomly chosen with a small probability, normally, $1/N$, where $N$ is the number of particles in the box. After the increment or decrement of the volume of the box, the particle positions are rescaled. The transition probability of acceptance of the changes in volume is
\begin{equation}
\frac{\rho_j}{\rho_i} = \left(\frac{V_j}{V_i}\right)^N e^{-\beta E_j+\beta E_i - P (V_j-V_i)} \ ,
\end{equation}
where $j$ is the new state, $V_i$ and $V_j$ are the volume accessible to the $N$ particles in the old and new states, respectively. The pressure $P$ is the input of the simulation. Again, if a random number uniformly distributed between $0$ and $1$ is lower than the transition probability, the movement is accepted. Otherwise it is rejected.

\end{appendix}

\bibliography{ref}

\begin{thebibliography}{46}%
\makeatletter
\providecommand \@ifxundefined [1]{%
 \@ifx{#1\undefined}
}%
\providecommand \@ifnum [1]{%
 \ifnum #1\expandafter \@firstoftwo
 \else \expandafter \@secondoftwo
 \fi
}%
\providecommand \@ifx [1]{%
 \ifx #1\expandafter \@firstoftwo
 \else \expandafter \@secondoftwo
 \fi
}%
\providecommand \natexlab [1]{#1}%
\providecommand \enquote  [1]{``#1''}%
\providecommand \bibnamefont  [1]{#1}%
\providecommand \bibfnamefont [1]{#1}%
\providecommand \citenamefont [1]{#1}%
\providecommand \href@noop [0]{\@secondoftwo}%
\providecommand \href [0]{\begingroup \@sanitize@url \@href}%
\providecommand \@href[1]{\@@startlink{#1}\@@href}%
\providecommand \@@href[1]{\endgroup#1\@@endlink}%
\providecommand \@sanitize@url [0]{\catcode `\\12\catcode `\$12\catcode
  `\&12\catcode `\#12\catcode `\^12\catcode `\_12\catcode `\%12\relax}%
\providecommand \@@startlink[1]{}%
\providecommand \@@endlink[0]{}%
\providecommand \url  [0]{\begingroup\@sanitize@url \@url }%
\providecommand \@url [1]{\endgroup\@href {#1}{\urlprefix }}%
\providecommand \urlprefix  [0]{URL }%
\providecommand \Eprint [0]{\href }%
\providecommand \doibase [0]{http://dx.doi.org/}%
\providecommand \selectlanguage [0]{\@gobble}%
\providecommand \bibinfo  [0]{\@secondoftwo}%
\providecommand \bibfield  [0]{\@secondoftwo}%
\providecommand \translation [1]{[#1]}%
\providecommand \BibitemOpen [0]{}%
\providecommand \bibitemStop [0]{}%
\providecommand \bibitemNoStop [0]{.\EOS\space}%
\providecommand \EOS [0]{\spacefactor3000\relax}%
\providecommand \BibitemShut  [1]{\csname bibitem#1\endcsname}%
\let\auto@bib@innerbib\@empty
\bibitem [{\citenamefont {Israelachvili}(1992)}]{Israelachvili}%
  \BibitemOpen
  \bibfield  {author} {\bibinfo {author} {\bibfnamefont {J.~N.}\ \bibnamefont
  {Israelachvili}},\ }\href@noop {} {\emph {\bibinfo {title} {Intermolecular
  and Surface Forces}}}\ (\bibinfo  {publisher} {Academic, New York},\ \bibinfo
  {year} {1992})\BibitemShut {NoStop}%
\bibitem [{\citenamefont {Evans}\ and\ \citenamefont
  {Wennerstrom}(1999)}]{Evans}%
  \BibitemOpen
  \bibfield  {author} {\bibinfo {author} {\bibfnamefont {D.~F.}\ \bibnamefont
  {Evans}}\ and\ \bibinfo {author} {\bibfnamefont {H.}~\bibnamefont
  {Wennerstrom}},\ }\href@noop {} {\emph {\bibinfo {title} {The Colloidal
  Domain}}}\ (\bibinfo  {publisher} {VCH, New York},\ \bibinfo {year}
  {1999})\BibitemShut {NoStop}%
\bibitem [{\citenamefont {Derjaguin}\ and\ \citenamefont
  {Landau}(1941)}]{DeLa41}%
  \BibitemOpen
  \bibfield  {author} {\bibinfo {author} {\bibfnamefont {B.~V.}\ \bibnamefont
  {Derjaguin}}\ and\ \bibinfo {author} {\bibfnamefont {L.}~\bibnamefont
  {Landau}},\ }\href@noop {} {\bibfield  {journal} {\bibinfo  {journal} {Acta
  Physicochimica (USSR)}\ }\textbf {\bibinfo {volume} {14}},\ \bibinfo {pages}
  {633} (\bibinfo {year} {1941})}\BibitemShut {NoStop}%
\bibitem [{\citenamefont {Verwey}\ and\ \citenamefont
  {Overbeek}(1948)}]{VeOv48}%
  \BibitemOpen
  \bibfield  {author} {\bibinfo {author} {\bibfnamefont {E.~J.~W.}\
  \bibnamefont {Verwey}}\ and\ \bibinfo {author} {\bibfnamefont {J.~T.~G.}\
  \bibnamefont {Overbeek}},\ }\href@noop {} {\emph {\bibinfo {title} {Theory of
  the Stability of Lyophobic Colloids}}}\ (\bibinfo  {publisher} {Elsevier,
  Amsterdam},\ \bibinfo {year} {1948})\BibitemShut {NoStop}%
\bibitem [{\citenamefont {Levin}(2002)}]{Levin}%
  \BibitemOpen
  \bibfield  {author} {\bibinfo {author} {\bibfnamefont {Y.}~\bibnamefont
  {Levin}},\ }\href@noop {} {\bibfield  {journal} {\bibinfo  {journal} {Rep.
  Prog. Phys.}\ }\textbf {\bibinfo {volume} {65}},\ \bibinfo {pages} {1577}
  (\bibinfo {year} {2002})}\BibitemShut {NoStop}%
\bibitem [{\citenamefont {Patey}(1980)}]{Pa80}%
  \BibitemOpen
  \bibfield  {author} {\bibinfo {author} {\bibfnamefont {G.~N.}\ \bibnamefont
  {Patey}},\ }\href@noop {} {\bibfield  {journal} {\bibinfo  {journal} {J.
  Chem. Phys.}\ }\textbf {\bibinfo {volume} {72}},\ \bibinfo {pages} {5763}
  (\bibinfo {year} {1980})}\BibitemShut {NoStop}%
\bibitem [{\citenamefont {Guldbrand}\ \emph {et~al.}(1984)\citenamefont
  {Guldbrand}, \citenamefont {Jonsson}, \citenamefont {Wennerstrom},\ and\
  \citenamefont {Linse}}]{GuJo84}%
  \BibitemOpen
  \bibfield  {author} {\bibinfo {author} {\bibfnamefont {L.}~\bibnamefont
  {Guldbrand}}, \bibinfo {author} {\bibfnamefont {B.}~\bibnamefont {Jonsson}},
  \bibinfo {author} {\bibfnamefont {H.}~\bibnamefont {Wennerstrom}}, \ and\
  \bibinfo {author} {\bibfnamefont {P.}~\bibnamefont {Linse}},\ }\href@noop {}
  {\bibfield  {journal} {\bibinfo  {journal} {J. Chem. Phys.}\ }\textbf
  {\bibinfo {volume} {80}},\ \bibinfo {pages} {2221} (\bibinfo {year}
  {1984})}\BibitemShut {NoStop}%
\bibitem [{\citenamefont {Stevens}\ and\ \citenamefont
  {Robbins}(1990)}]{StRo90}%
  \BibitemOpen
  \bibfield  {author} {\bibinfo {author} {\bibfnamefont {M.~J.}\ \bibnamefont
  {Stevens}}\ and\ \bibinfo {author} {\bibfnamefont {M.~O.}\ \bibnamefont
  {Robbins}},\ }\href@noop {} {\bibfield  {journal} {\bibinfo  {journal}
  {Europhys. Lett.}\ }\textbf {\bibinfo {volume} {12}},\ \bibinfo {pages} {81}
  (\bibinfo {year} {1990})}\BibitemShut {NoStop}%
\bibitem [{\citenamefont {Larsen}\ and\ \citenamefont {Grier}(1997)}]{LaGr97}%
  \BibitemOpen
  \bibfield  {author} {\bibinfo {author} {\bibfnamefont {A.~E.}\ \bibnamefont
  {Larsen}}\ and\ \bibinfo {author} {\bibfnamefont {D.~G.}\ \bibnamefont
  {Grier}},\ }\href@noop {} {\bibfield  {journal} {\bibinfo  {journal}
  {Nature}\ }\textbf {\bibinfo {volume} {385}},\ \bibinfo {pages} {230}
  (\bibinfo {year} {1997})}\BibitemShut {NoStop}%
\bibitem [{\citenamefont {Solis}\ and\ \citenamefont {{de la
  Cruz}}(1999)}]{SoDe99}%
  \BibitemOpen
  \bibfield  {author} {\bibinfo {author} {\bibfnamefont {F.}~\bibnamefont
  {Solis}}\ and\ \bibinfo {author} {\bibfnamefont {M.}~\bibnamefont {{de la
  Cruz}}},\ }\href@noop {} {\bibfield  {journal} {\bibinfo  {journal} {Phys.
  Rev. E}\ }\textbf {\bibinfo {volume} {60}},\ \bibinfo {pages} {4496}
  (\bibinfo {year} {1999})}\BibitemShut {NoStop}%
\bibitem [{\citenamefont {Levin}, \citenamefont {Arenzon},\ and\ \citenamefont
  {Stilck}(1999)}]{LeAr99}%
  \BibitemOpen
  \bibfield  {author} {\bibinfo {author} {\bibfnamefont {Y.}~\bibnamefont
  {Levin}}, \bibinfo {author} {\bibfnamefont {J.~J.}\ \bibnamefont {Arenzon}},
  \ and\ \bibinfo {author} {\bibfnamefont {J.~F.}\ \bibnamefont {Stilck}},\
  }\href@noop {} {\bibfield  {journal} {\bibinfo  {journal} {Phys. Rev. Lett.}\
  }\textbf {\bibinfo {volume} {83}},\ \bibinfo {pages} {2680} (\bibinfo {year}
  {1999})}\BibitemShut {NoStop}%
\bibitem [{\citenamefont {Linse}\ and\ \citenamefont
  {Lobaskin}(1999)}]{LiLo99}%
  \BibitemOpen
  \bibfield  {author} {\bibinfo {author} {\bibfnamefont {P.}~\bibnamefont
  {Linse}}\ and\ \bibinfo {author} {\bibfnamefont {V.}~\bibnamefont
  {Lobaskin}},\ }\href@noop {} {\bibfield  {journal} {\bibinfo  {journal}
  {Phys. Rev. Lett.}\ }\textbf {\bibinfo {volume} {83}},\ \bibinfo {pages}
  {4208} (\bibinfo {year} {1999})}\BibitemShut {NoStop}%
\bibitem [{\citenamefont {Butler}\ \emph {et~al.}(2003)\citenamefont {Butler},
  \citenamefont {Angelini}, \citenamefont {Tang},\ and\ \citenamefont
  {Wong}}]{BuAn03}%
  \BibitemOpen
  \bibfield  {author} {\bibinfo {author} {\bibfnamefont {J.~C.}\ \bibnamefont
  {Butler}}, \bibinfo {author} {\bibfnamefont {T.}~\bibnamefont {Angelini}},
  \bibinfo {author} {\bibfnamefont {J.~X.}\ \bibnamefont {Tang}}, \ and\
  \bibinfo {author} {\bibfnamefont {G.~C.~L.}\ \bibnamefont {Wong}},\
  }\href@noop {} {\bibfield  {journal} {\bibinfo  {journal} {Phys. Rev. Lett.}\
  }\textbf {\bibinfo {volume} {91}},\ \bibinfo {pages} {028301} (\bibinfo
  {year} {2003})}\BibitemShut {NoStop}%
\bibitem [{\citenamefont {Samaj}\ and\ \citenamefont {Trizac}(2011)}]{SaTr11}%
  \BibitemOpen
  \bibfield  {author} {\bibinfo {author} {\bibfnamefont {L.}~\bibnamefont
  {Samaj}}\ and\ \bibinfo {author} {\bibfnamefont {E.}~\bibnamefont {Trizac}},\
  }\href@noop {} {\bibfield  {journal} {\bibinfo  {journal} {Phys. Rev. Lett.}\
  }\textbf {\bibinfo {volume} {106}},\ \bibinfo {pages} {078301} (\bibinfo
  {year} {2011})}\BibitemShut {NoStop}%
\bibitem [{\citenamefont {Zhou}(2013)}]{Zh13}%
  \BibitemOpen
  \bibfield  {author} {\bibinfo {author} {\bibfnamefont {S.}~\bibnamefont
  {Zhou}},\ }\href@noop {} {\bibfield  {journal} {\bibinfo  {journal}
  {Langmuir}\ }\textbf {\bibinfo {volume} {29}},\ \bibinfo {pages} {12490}
  (\bibinfo {year} {2013})}\BibitemShut {NoStop}%
\bibitem [{\citenamefont {Bostr\"om}, \citenamefont {Williams},\ and\
  \citenamefont {Ninham}(2001)}]{BoWi01}%
  \BibitemOpen
  \bibfield  {author} {\bibinfo {author} {\bibfnamefont {M.}~\bibnamefont
  {Bostr\"om}}, \bibinfo {author} {\bibfnamefont {D.~R.~M.}\ \bibnamefont
  {Williams}}, \ and\ \bibinfo {author} {\bibfnamefont {B.~W.}\ \bibnamefont
  {Ninham}},\ }\href@noop {} {\bibfield  {journal} {\bibinfo  {journal}
  {Langmuir}\ }\textbf {\bibinfo {volume} {17}},\ \bibinfo {pages} {4475}
  (\bibinfo {year} {2001})}\BibitemShut {NoStop}%
\bibitem [{\citenamefont {Lopez-Leon}\ \emph {et~al.}(2003)\citenamefont
  {Lopez-Leon}, \citenamefont {Jodar-Reyes}, \citenamefont {Bastos-Gonzalez},\
  and\ \citenamefont {Ortega-Vinuesa}}]{LoJo03}%
  \BibitemOpen
  \bibfield  {author} {\bibinfo {author} {\bibfnamefont {T.}~\bibnamefont
  {Lopez-Leon}}, \bibinfo {author} {\bibfnamefont {A.~B.}\ \bibnamefont
  {Jodar-Reyes}}, \bibinfo {author} {\bibfnamefont {D.}~\bibnamefont
  {Bastos-Gonzalez}}, \ and\ \bibinfo {author} {\bibfnamefont {J.~L.}\
  \bibnamefont {Ortega-Vinuesa}},\ }\href@noop {} {\bibfield  {journal}
  {\bibinfo  {journal} {J. Phys. Chem. B}\ }\textbf {\bibinfo {volume} {107}},\
  \bibinfo {pages} {5696} (\bibinfo {year} {2003})}\BibitemShut {NoStop}%
\bibitem [{\citenamefont {Lopez-Leon}\ \emph {et~al.}(2008)\citenamefont
  {Lopez-Leon}, \citenamefont {Santander-Ortega}, \citenamefont
  {Ortega-Vinuesa},\ and\ \citenamefont {Bastos-Gonzalez}}]{LoSa08}%
  \BibitemOpen
  \bibfield  {author} {\bibinfo {author} {\bibfnamefont {T.}~\bibnamefont
  {Lopez-Leon}}, \bibinfo {author} {\bibfnamefont {M.~J.}\ \bibnamefont
  {Santander-Ortega}}, \bibinfo {author} {\bibfnamefont {J.~L.}\ \bibnamefont
  {Ortega-Vinuesa}}, \ and\ \bibinfo {author} {\bibfnamefont {D.}~\bibnamefont
  {Bastos-Gonzalez}},\ }\href@noop {} {\bibfield  {journal} {\bibinfo
  {journal} {J. Phys. Chem. C}\ }\textbf {\bibinfo {volume} {112}},\ \bibinfo
  {pages} {16060} (\bibinfo {year} {2008})}\BibitemShut {NoStop}%
\bibitem [{\citenamefont {Peula-Garcia}, \citenamefont {Ortega-Vinuesa},\ and\
  \citenamefont {Bastos-Gonzalez}(2010)}]{PeOr10}%
  \BibitemOpen
  \bibfield  {author} {\bibinfo {author} {\bibfnamefont {J.~M.}\ \bibnamefont
  {Peula-Garcia}}, \bibinfo {author} {\bibfnamefont {J.~L.}\ \bibnamefont
  {Ortega-Vinuesa}}, \ and\ \bibinfo {author} {\bibfnamefont {D.}~\bibnamefont
  {Bastos-Gonzalez}},\ }\href@noop {} {\bibfield  {journal} {\bibinfo
  {journal} {J. Phys. Chem. C}\ }\textbf {\bibinfo {volume} {114}},\ \bibinfo
  {pages} {11133} (\bibinfo {year} {2010})}\BibitemShut {NoStop}%
\bibitem [{\citenamefont {{dos Santos}}\ and\ \citenamefont
  {Levin}(2011)}]{DoLe11}%
  \BibitemOpen
  \bibfield  {author} {\bibinfo {author} {\bibfnamefont {A.~P.}\ \bibnamefont
  {{dos Santos}}}\ and\ \bibinfo {author} {\bibfnamefont {Y.}~\bibnamefont
  {Levin}},\ }\href@noop {} {\bibfield  {journal} {\bibinfo  {journal} {Phys.
  Rev. Lett.}\ }\textbf {\bibinfo {volume} {106}},\ \bibinfo {pages} {167801}
  (\bibinfo {year} {2011})}\BibitemShut {NoStop}%
\bibitem [{\citenamefont {Miklavic}\ \emph {et~al.}(1994)\citenamefont
  {Miklavic}, \citenamefont {Chan}, \citenamefont {White},\ and\ \citenamefont
  {Healy}}]{MiCh94}%
  \BibitemOpen
  \bibfield  {author} {\bibinfo {author} {\bibfnamefont {S.~J.}\ \bibnamefont
  {Miklavic}}, \bibinfo {author} {\bibfnamefont {D.~Y.~C.}\ \bibnamefont
  {Chan}}, \bibinfo {author} {\bibfnamefont {L.~R.}\ \bibnamefont {White}}, \
  and\ \bibinfo {author} {\bibfnamefont {T.~W.}\ \bibnamefont {Healy}},\
  }\href@noop {} {\bibfield  {journal} {\bibinfo  {journal} {J. Phys. Chem.}\
  }\textbf {\bibinfo {volume} {98}},\ \bibinfo {pages} {9022} (\bibinfo {year}
  {1994})}\BibitemShut {NoStop}%
\bibitem [{\citenamefont {Miklavic}(1995)}]{Mi95}%
  \BibitemOpen
  \bibfield  {author} {\bibinfo {author} {\bibfnamefont {S.~J.}\ \bibnamefont
  {Miklavic}},\ }\href@noop {} {\bibfield  {journal} {\bibinfo  {journal} {J.
  Chem. Phys.}\ }\textbf {\bibinfo {volume} {103}},\ \bibinfo {pages} {4794}
  (\bibinfo {year} {1995})}\BibitemShut {NoStop}%
\bibitem [{\citenamefont {Zhang}\ \emph {et~al.}(2005)\citenamefont {Zhang},
  \citenamefont {Yoon}, \citenamefont {Mao},\ and\ \citenamefont
  {Ducker}}]{ZhYo05}%
  \BibitemOpen
  \bibfield  {author} {\bibinfo {author} {\bibfnamefont {J.}~\bibnamefont
  {Zhang}}, \bibinfo {author} {\bibfnamefont {R.~H.}\ \bibnamefont {Yoon}},
  \bibinfo {author} {\bibfnamefont {M.}~\bibnamefont {Mao}}, \ and\ \bibinfo
  {author} {\bibfnamefont {W.~A.}\ \bibnamefont {Ducker}},\ }\href@noop {}
  {\bibfield  {journal} {\bibinfo  {journal} {Langmuir}\ }\textbf {\bibinfo
  {volume} {21}},\ \bibinfo {pages} {5831} (\bibinfo {year}
  {2005})}\BibitemShut {NoStop}%
\bibitem [{\citenamefont {Meyer}\ \emph {et~al.}(2005)\citenamefont {Meyer},
  \citenamefont {Lin}, \citenamefont {Hassenkam}, \citenamefont {Oroudjev},\
  and\ \citenamefont {Israelachvili}}]{MeLi05}%
  \BibitemOpen
  \bibfield  {author} {\bibinfo {author} {\bibfnamefont {E.~E.}\ \bibnamefont
  {Meyer}}, \bibinfo {author} {\bibfnamefont {Q.}~\bibnamefont {Lin}}, \bibinfo
  {author} {\bibfnamefont {T.}~\bibnamefont {Hassenkam}}, \bibinfo {author}
  {\bibfnamefont {E.}~\bibnamefont {Oroudjev}}, \ and\ \bibinfo {author}
  {\bibfnamefont {J.~N.}\ \bibnamefont {Israelachvili}},\ }\href@noop {}
  {\bibfield  {journal} {\bibinfo  {journal} {Proc. Natl. Acad. Sci. U.S.A}\
  }\textbf {\bibinfo {volume} {102}},\ \bibinfo {pages} {6839} (\bibinfo {year}
  {2005})}\BibitemShut {NoStop}%
\bibitem [{\citenamefont {Ben-Yaakov}\ \emph {et~al.}(2007)\citenamefont
  {Ben-Yaakov}, \citenamefont {Burak}, \citenamefont {Andelman},\ and\
  \citenamefont {Safran}}]{BeBu07}%
  \BibitemOpen
  \bibfield  {author} {\bibinfo {author} {\bibfnamefont {D.}~\bibnamefont
  {Ben-Yaakov}}, \bibinfo {author} {\bibfnamefont {Y.}~\bibnamefont {Burak}},
  \bibinfo {author} {\bibfnamefont {D.}~\bibnamefont {Andelman}}, \ and\
  \bibinfo {author} {\bibfnamefont {S.~A.}\ \bibnamefont {Safran}},\
  }\href@noop {} {\bibfield  {journal} {\bibinfo  {journal} {Europhys Lett}\
  }\textbf {\bibinfo {volume} {79}},\ \bibinfo {pages} {48002} (\bibinfo {year}
  {2007})}\BibitemShut {NoStop}%
\bibitem [{\citenamefont {Tsao}, \citenamefont {Evans},\ and\ \citenamefont
  {Wennerstrom}(1993)}]{Tsao}%
  \BibitemOpen
  \bibfield  {author} {\bibinfo {author} {\bibfnamefont {Y.~H.}\ \bibnamefont
  {Tsao}}, \bibinfo {author} {\bibfnamefont {D.~F.}\ \bibnamefont {Evans}}, \
  and\ \bibinfo {author} {\bibfnamefont {H.}~\bibnamefont {Wennerstrom}},\
  }\href@noop {} {\bibfield  {journal} {\bibinfo  {journal} {Science}\ }\textbf
  {\bibinfo {volume} {262}},\ \bibinfo {pages} {547} (\bibinfo {year}
  {1993})}\BibitemShut {NoStop}%
\bibitem [{\citenamefont {Brewster}, \citenamefont {Pincus},\ and\
  \citenamefont {Safran}(2008)}]{Brewster}%
  \BibitemOpen
  \bibfield  {author} {\bibinfo {author} {\bibfnamefont {R.}~\bibnamefont
  {Brewster}}, \bibinfo {author} {\bibfnamefont {P.~A.}\ \bibnamefont
  {Pincus}}, \ and\ \bibinfo {author} {\bibfnamefont {S.~A.}\ \bibnamefont
  {Safran}},\ }\href@noop {} {\bibfield  {journal} {\bibinfo  {journal} {Phys.
  Rev. Lett}\ }\textbf {\bibinfo {volume} {101}},\ \bibinfo {pages} {128101}
  (\bibinfo {year} {2008})}\BibitemShut {NoStop}%
\bibitem [{\citenamefont {Jho}\ \emph {et~al.}(2011)\citenamefont {Jho},
  \citenamefont {Brewster}, \citenamefont {Safran},\ and\ \citenamefont
  {Pincus}}]{Jho}%
  \BibitemOpen
  \bibfield  {author} {\bibinfo {author} {\bibfnamefont {Y.~S.}\ \bibnamefont
  {Jho}}, \bibinfo {author} {\bibfnamefont {R.}~\bibnamefont {Brewster}},
  \bibinfo {author} {\bibfnamefont {S.~A.}\ \bibnamefont {Safran}}, \ and\
  \bibinfo {author} {\bibfnamefont {P.~A.}\ \bibnamefont {Pincus}},\
  }\href@noop {} {\bibfield  {journal} {\bibinfo  {journal} {Langmuir}\
  }\textbf {\bibinfo {volume} {27}},\ \bibinfo {pages} {4439} (\bibinfo {year}
  {2011})}\BibitemShut {NoStop}%
\bibitem [{\citenamefont {Silbert}\ \emph {et~al.}(2012)\citenamefont
  {Silbert}, \citenamefont {Ben-Yaakov}, \citenamefont {Dror}, \citenamefont
  {Perkin}, \citenamefont {Kampf},\ and\ \citenamefont {Klein}}]{Silbert}%
  \BibitemOpen
  \bibfield  {author} {\bibinfo {author} {\bibfnamefont {G.}~\bibnamefont
  {Silbert}}, \bibinfo {author} {\bibfnamefont {D.}~\bibnamefont {Ben-Yaakov}},
  \bibinfo {author} {\bibfnamefont {Y.}~\bibnamefont {Dror}}, \bibinfo {author}
  {\bibfnamefont {S.}~\bibnamefont {Perkin}}, \bibinfo {author} {\bibfnamefont
  {N.}~\bibnamefont {Kampf}}, \ and\ \bibinfo {author} {\bibfnamefont
  {J.}~\bibnamefont {Klein}},\ }\href@noop {} {\bibfield  {journal} {\bibinfo
  {journal} {Phys. Rev. Lett.}\ }\textbf {\bibinfo {volume} {109}},\ \bibinfo
  {pages} {168305} (\bibinfo {year} {2012})}\BibitemShut {NoStop}%
\bibitem [{\citenamefont {Naji}\ and\ \citenamefont {Podgornik}(2005)}]{Naji}%
  \BibitemOpen
  \bibfield  {author} {\bibinfo {author} {\bibfnamefont {A.}~\bibnamefont
  {Naji}}\ and\ \bibinfo {author} {\bibfnamefont {R.}~\bibnamefont
  {Podgornik}},\ }\href@noop {} {\bibfield  {journal} {\bibinfo  {journal}
  {Phys. Rev. E.}\ }\textbf {\bibinfo {volume} {72}},\ \bibinfo {pages}
  {041402} (\bibinfo {year} {2005})}\BibitemShut {NoStop}%
\bibitem [{\citenamefont {Ben-Yaakov}, \citenamefont {Andelman},\ and\
  \citenamefont {Diamant}(2013)}]{Andelman}%
  \BibitemOpen
  \bibfield  {author} {\bibinfo {author} {\bibfnamefont {D.}~\bibnamefont
  {Ben-Yaakov}}, \bibinfo {author} {\bibfnamefont {D.}~\bibnamefont
  {Andelman}}, \ and\ \bibinfo {author} {\bibfnamefont {H.}~\bibnamefont
  {Diamant}},\ }\href@noop {} {\bibfield  {journal} {\bibinfo  {journal} {Phys.
  Rev. E}\ }\textbf {\bibinfo {volume} {87}},\ \bibinfo {pages} {022402}
  (\bibinfo {year} {2013})}\BibitemShut {NoStop}%
\bibitem [{\citenamefont {Sarabadani}\ \emph {et~al.}(2010)\citenamefont
  {Sarabadani}, \citenamefont {Naji}, \citenamefont {Dean}, \citenamefont
  {Horgan},\ and\ \citenamefont {Podgornik}}]{Sarabadani}%
  \BibitemOpen
  \bibfield  {author} {\bibinfo {author} {\bibfnamefont {J.}~\bibnamefont
  {Sarabadani}}, \bibinfo {author} {\bibfnamefont {A.}~\bibnamefont {Naji}},
  \bibinfo {author} {\bibfnamefont {D.~S.}\ \bibnamefont {Dean}}, \bibinfo
  {author} {\bibfnamefont {R.~R.}\ \bibnamefont {Horgan}}, \ and\ \bibinfo
  {author} {\bibfnamefont {R.}~\bibnamefont {Podgornik}},\ }\href@noop {}
  {\bibfield  {journal} {\bibinfo  {journal} {J. Chem. Phys.}\ }\textbf
  {\bibinfo {volume} {133}},\ \bibinfo {pages} {174702} (\bibinfo {year}
  {2010})}\BibitemShut {NoStop}%
\bibitem [{\citenamefont {Valleau}\ and\ \citenamefont
  {Cohen}(1980)}]{Valleau}%
  \BibitemOpen
  \bibfield  {author} {\bibinfo {author} {\bibfnamefont {J.~P.}\ \bibnamefont
  {Valleau}}\ and\ \bibinfo {author} {\bibfnamefont {L.~K.}\ \bibnamefont
  {Cohen}},\ }\href@noop {} {\bibfield  {journal} {\bibinfo  {journal} {J.
  Chem. Phys.}\ }\textbf {\bibinfo {volume} {72}},\ \bibinfo {pages} {5935}
  (\bibinfo {year} {1980})}\BibitemShut {NoStop}%
\bibitem [{\citenamefont {Frenkel}\ and\ \citenamefont
  {Smith}(1996)}]{Frenkel}%
  \BibitemOpen
  \bibfield  {author} {\bibinfo {author} {\bibfnamefont {D.}~\bibnamefont
  {Frenkel}}\ and\ \bibinfo {author} {\bibfnamefont {B.}~\bibnamefont
  {Smith}},\ }\href@noop {} {\emph {\bibinfo {title} {Understanding Molecular
  Simulation}}}\ (\bibinfo  {publisher} {Academic, New York},\ \bibinfo {year}
  {1996})\BibitemShut {NoStop}%
\bibitem [{\citenamefont {Allen}\ and\ \citenamefont
  {Tildesley}(1987)}]{Allen}%
  \BibitemOpen
  \bibfield  {author} {\bibinfo {author} {\bibfnamefont {M.~P.}\ \bibnamefont
  {Allen}}\ and\ \bibinfo {author} {\bibfnamefont {D.~J.}\ \bibnamefont
  {Tildesley}},\ }\href@noop {} {\emph {\bibinfo {title} {Computer Simulations
  of Liquids}}}\ (\bibinfo  {publisher} {Oxford: Oxford University Press},\
  \bibinfo {year} {1987})\BibitemShut {NoStop}%
\bibitem [{\citenamefont {Ho}, \citenamefont {Tsao},\ and\ \citenamefont
  {Sheng}(2003)}]{Ching}%
  \BibitemOpen
  \bibfield  {author} {\bibinfo {author} {\bibfnamefont {C.~H.}\ \bibnamefont
  {Ho}}, \bibinfo {author} {\bibfnamefont {H.~K.}\ \bibnamefont {Tsao}}, \ and\
  \bibinfo {author} {\bibfnamefont {Y.~J.}\ \bibnamefont {Sheng}},\ }\href@noop
  {} {\bibfield  {journal} {\bibinfo  {journal} {J. Chem. Phys.}\ }\textbf
  {\bibinfo {volume} {119}},\ \bibinfo {pages} {2369} (\bibinfo {year}
  {2003})}\BibitemShut {NoStop}%
\bibitem [{\citenamefont {Levin}\ and\ \citenamefont {Fisher}(1996)}]{Fisher}%
  \BibitemOpen
  \bibfield  {author} {\bibinfo {author} {\bibfnamefont {Y.}~\bibnamefont
  {Levin}}\ and\ \bibinfo {author} {\bibfnamefont {M.~E.}\ \bibnamefont
  {Fisher}},\ }\href@noop {} {\bibfield  {journal} {\bibinfo  {journal}
  {Physica A}\ }\textbf {\bibinfo {volume} {225}},\ \bibinfo {pages} {164}
  (\bibinfo {year} {1996})}\BibitemShut {NoStop}%
\bibitem [{\citenamefont {Lebowitz}\ and\ \citenamefont
  {Waisman}(1972{\natexlab{a}})}]{LeWa72a}%
  \BibitemOpen
  \bibfield  {author} {\bibinfo {author} {\bibfnamefont {J.~L.}\ \bibnamefont
  {Lebowitz}}\ and\ \bibinfo {author} {\bibfnamefont {E.}~\bibnamefont
  {Waisman}},\ }\href@noop {} {\bibfield  {journal} {\bibinfo  {journal} {J.
  Chem. Phys.}\ }\textbf {\bibinfo {volume} {56}},\ \bibinfo {pages} {3086}
  (\bibinfo {year} {1972}{\natexlab{a}})}\BibitemShut {NoStop}%
\bibitem [{\citenamefont {Lebowitz}\ and\ \citenamefont
  {Waisman}(1972{\natexlab{b}})}]{LeWa72b}%
  \BibitemOpen
  \bibfield  {author} {\bibinfo {author} {\bibfnamefont {J.~L.}\ \bibnamefont
  {Lebowitz}}\ and\ \bibinfo {author} {\bibfnamefont {E.}~\bibnamefont
  {Waisman}},\ }\href@noop {} {\bibfield  {journal} {\bibinfo  {journal} {J.
  Chem. Phys.}\ }\textbf {\bibinfo {volume} {56}},\ \bibinfo {pages} {3093}
  (\bibinfo {year} {1972}{\natexlab{b}})}\BibitemShut {NoStop}%
\bibitem [{\citenamefont {Wu}\ \emph {et~al.}(1999)\citenamefont {Wu},
  \citenamefont {Bratko}, \citenamefont {Blanch},\ and\ \citenamefont
  {Prausnitz}}]{Wu}%
  \BibitemOpen
  \bibfield  {author} {\bibinfo {author} {\bibfnamefont {J.~Z.}\ \bibnamefont
  {Wu}}, \bibinfo {author} {\bibfnamefont {D.}~\bibnamefont {Bratko}}, \bibinfo
  {author} {\bibfnamefont {H.~W.}\ \bibnamefont {Blanch}}, \ and\ \bibinfo
  {author} {\bibfnamefont {J.~M.}\ \bibnamefont {Prausnitz}},\ }\href@noop {}
  {\bibfield  {journal} {\bibinfo  {journal} {J. Chem. Phys.}\ }\textbf
  {\bibinfo {volume} {111}},\ \bibinfo {pages} {7084} (\bibinfo {year}
  {1999})}\BibitemShut {NoStop}%
\bibitem [{\citenamefont {Colla}, \citenamefont {dos Santos},\ and\
  \citenamefont {Levin}(2012)}]{Alexander}%
  \BibitemOpen
  \bibfield  {author} {\bibinfo {author} {\bibfnamefont {T.~E.}\ \bibnamefont
  {Colla}}, \bibinfo {author} {\bibfnamefont {A.~P.}\ \bibnamefont {dos
  Santos}}, \ and\ \bibinfo {author} {\bibfnamefont {Y.}~\bibnamefont
  {Levin}},\ }\href@noop {} {\bibfield  {journal} {\bibinfo  {journal} {J.
  Chem. Phys.}\ }\textbf {\bibinfo {volume} {136}},\ \bibinfo {pages} {194103}
  (\bibinfo {year} {2012})}\BibitemShut {NoStop}%
\bibitem [{\citenamefont {Yeh}\ and\ \citenamefont {Berkowitz}(1981)}]{Yeh}%
  \BibitemOpen
  \bibfield  {author} {\bibinfo {author} {\bibfnamefont {I.~C.}\ \bibnamefont
  {Yeh}}\ and\ \bibinfo {author} {\bibfnamefont {M.~L.}\ \bibnamefont
  {Berkowitz}},\ }\href@noop {} {\bibfield  {journal} {\bibinfo  {journal} {J.
  Chem. Phys.}\ }\textbf {\bibinfo {volume} {111}},\ \bibinfo {pages} {475}
  (\bibinfo {year} {1981})}\BibitemShut {NoStop}%
\bibitem [{\citenamefont {Tamashiro}, \citenamefont {Levin},\ and\
  \citenamefont {Barbosa}(1998)}]{TaLe98}%
  \BibitemOpen
  \bibfield  {author} {\bibinfo {author} {\bibfnamefont {M.~N.}\ \bibnamefont
  {Tamashiro}}, \bibinfo {author} {\bibfnamefont {Y.}~\bibnamefont {Levin}}, \
  and\ \bibinfo {author} {\bibfnamefont {M.~C.}\ \bibnamefont {Barbosa}},\
  }\href@noop {} {\bibfield  {journal} {\bibinfo  {journal} {Eur. Phys. J. B}\
  }\textbf {\bibinfo {volume} {1}},\ \bibinfo {pages} {337} (\bibinfo {year}
  {1998})}\BibitemShut {NoStop}%
\bibitem [{\citenamefont {Henderson}, \citenamefont {Blum},\ and\ \citenamefont
  {Lebowitz}(1979)}]{HeBl79}%
  \BibitemOpen
  \bibfield  {author} {\bibinfo {author} {\bibfnamefont {D.}~\bibnamefont
  {Henderson}}, \bibinfo {author} {\bibfnamefont {L.}~\bibnamefont {Blum}}, \
  and\ \bibinfo {author} {\bibfnamefont {J.~L.}\ \bibnamefont {Lebowitz}},\
  }\href@noop {} {\bibfield  {journal} {\bibinfo  {journal} {J. Electroanal.
  Chem.}\ }\textbf {\bibinfo {volume} {102}},\ \bibinfo {pages} {315} (\bibinfo
  {year} {1979})}\BibitemShut {NoStop}%
\bibitem [{\citenamefont {Blum}\ and\ \citenamefont
  {Henderson}(1981)}]{BlHe81}%
  \BibitemOpen
  \bibfield  {author} {\bibinfo {author} {\bibfnamefont {L.}~\bibnamefont
  {Blum}}\ and\ \bibinfo {author} {\bibfnamefont {D.}~\bibnamefont
  {Henderson}},\ }\href@noop {} {\bibfield  {journal} {\bibinfo  {journal} {J.
  Chem. Phys.}\ }\textbf {\bibinfo {volume} {74}},\ \bibinfo {pages} {1902}
  (\bibinfo {year} {1981})}\BibitemShut {NoStop}%
\bibitem [{\citenamefont {Carnie}\ and\ \citenamefont {Chan}(1981)}]{CaCh81}%
  \BibitemOpen
  \bibfield  {author} {\bibinfo {author} {\bibfnamefont {S.~L.}\ \bibnamefont
  {Carnie}}\ and\ \bibinfo {author} {\bibfnamefont {D.~Y.~C.}\ \bibnamefont
  {Chan}},\ }\href@noop {} {\bibfield  {journal} {\bibinfo  {journal} {J. Chem.
  Phys.}\ }\textbf {\bibinfo {volume} {74}},\ \bibinfo {pages} {1293} (\bibinfo
  {year} {1981})}\BibitemShut {NoStop}%
\end{thebibliography}%

\end{document}